# A Review on The Use of Deep Learning in Android Malware Detection

Abdelmonim Naway [1], Yuancheng LI [1]

[1] North China Electric Power University, School of Control and Computer Engineering, 2 Beinong Road, Champing District, Beijing, China,102206

{abdelmonim, yuanchengli}@ncepu.edu.cn

*Abstract—* Android is the predominant mobile operating system for the past few years. The prevalence of devices that can be powered by Android magnetized not merely application developers but also malware developers with criminal intention to design and spread malicious applications that can affect the normal work of Android phones and tablets, steal personal information and credential data, or even worse lock the phone and ask for ransom. Researchers persistently devise countermeasures strategies to fight back malware. One of these strategies applied in the past five years is the use of deep learning methods in Android malware detection. This necessitates a review to inspect the accomplished work in order to know where the endeavors have been established, identify unresolved problems, and motivate future research directions. In this work, an extensive survey of static analysis, dynamic analysis, and hybrid analysis that utilized deep learning methods are reviewed with an elaborated discussion on their key concepts, contributions, and limitations.

*Keywords—* Android Malware Detection, Static Analysis, Dynamic Analysis, Hybrid Analysis, Deep Learning

## I. INTRODUCTION

The accelerated expansion of Android malware has designated an immense obstacle in front of malware analysts. The researchers have constantly suggested defenses and designing novel methods to fight malware attacks. It is crucial for malware detection techniques to meet the menace of malware. In pursuing independent learning of malware identification and lowering human expert engagement, deep learning has been introduced into Android malware detection. Deep learning is a branch of machine learning that depends on studying various levels of representations, analogous to a ranking of features or notions, where top-level notions are determined from lower-level ones, and similar lower- level notions could assist in determining numerous top-level notions. Deep learning is a component of a large class of machine learning techniques that relied on learning representations. An observation (for instance an image) can be described by different means (e.g., a vector of pixels), yet some descriptions make it simpler to study tasks of concern from examples, and research in this field seeks to determine what makes better descriptions and how to study them. [40]. Techniques applied before deep learning, such as drawing separate levels of features from malicious samples for classification, which cannot mirror the overarching attributes of malware. Additionally, classification built on diverse varieties cast doubt on dimensions, time and computation resources. The utilization of deep learning for malware classification offers a means of building scalable machine learning models, which may handle any measure of data, without expending consistently of resources such as memory. Deep learning marks malware depend on the general pattern, which directs the distinguishing of a variety of malware attacks and their variations. Furthermore, deep learning conducts a profound classification and improves its accuracy because deep learning identifies more features than conventional machine learning methods by passing through many levels of features extraction. This enables deep learning models to acquire a new pattern of malware after the basic training phase.

Recently there were issues raised about machine learning and deep learning security as machine learning algorithms have been devised under the presumption that training and test data pursuing the equivalent basic probability distribution, which causes them to be exposed to skillfully-constructed attacks infringing this hypothesis. Deep learning methods can be substantially influenced by adversarial attacks using the experience of the learning algorithm to avoid detection, or infuse harming instances into the training data to deceive the learning algorithm and hence create incorrect classification results [39]. In the view of these issues and after five years of using deep learning in Android malware detection, there is a need for a comprehensive review of the current state-of-the-art about what has been studied, recognize where the concentrate has been established and specifies the direction of needed future research. In this work, we give a thorough review of the use of deep learning in Android malware analysis with respect to analysis type. Then we analyze and report on existing literature founded on specific criteria. Finally, we recognize open issues and suggest future research directions. In summary, this work provides the following contribution:

- To the best of our knowledge, this is the first review on the use of deep learning techniques in Android malware detection
- To present background information on Android malware analysis types with elaborated dissection of their strengths and weaknesses
- To organize an inclusive review of the work accomplished in Android malware analysis using deep learning with static analysis, dynamic analysis and hybrid analysis
- To determine the limitations of current approaches, list open issues, and recommend potential future works

## II. BACKGROUND IN ANDROID AND ANDROID MALWARE DETECTION

### 1. Android Application Components

Android applications are composed in Java programming language and can likewise be composed in C++. They are gathered and bundled in an APK (Android package). Every application keeps running in a different process and is made up of a required XML descriptor document called AndroidManifest.xml. The Android Manifest document contains every information required by the Android framework about the application. The AndroidManifest.xml allows determining the packages, APIs, libraries required by the application, permission imposed and asked for by the application, etc... Applications comprised of four parts: Activity, Service, Content Provider, and Broadcast Receiver These parts carry information through messages called Intents [30].

### 2. Android Malware Detection Techniques

Mobile malware requires sophisticated analysis. One critical point of mobile phones is that they are a sensor-based event system, which permits malware to respond to approaching SMS, position changes and so forth, increasing the sophistication of automated malware-analysis techniques. Moreover, the apps can use services and activities and integrate varied programming languages (e.g. Java and c++) in one application [33]. Resting on the features employed to classify an application, the analysis is organized as static analysis, dynamic analysis and hybrid analysis that combine static and dynamic analysis. Each one of these analyses has its strengths and weaknesses. In the following section, these techniques are discussed with their benefits and limitations.

### 3. Static Analysis

The static analysis screens parts of the application without really executing them. This technique incorporates Signature-based, Permission-based and Component-based analysis. The Signature-based strategy draws features and makes distinctive signs to identify specific malware. Hence, it falls short to recognize the variation or unidentified malware. The Permission-based strategy recognizes permission requests to distinguish malware. The Component-based techniques decompile the APP to draw and inspect the definition and byte code connections of significant components (i.e. activities, services, etc.), to identify the exposures [28] [31]. The principal drawbacks of static analysis are the missing of real execution paths and suitable execution conditions. Additionally, there exist problems in the occurrence of code obfuscation and dynamic code loading [29].

### 4. Dynamic Analysis

The dynamic analysis technique includes the execution of the application on either a virtual machine or a physical device. Amid the examination, the behavior of the application is watched and can be dissected. The dynamic analysis results in a less abstract perspective of application than static analysis. The code paths executed during runtime are a subset of every single accessible path. The principle objective of the analysis is to achieve high code inclusion since every feasible event ought to be activated to watch any possible malicious behavior [32]. The principal drawbacks of dynamic analysis are: dynamic analysis requires such considerable resources with respect to static analysis, which obstructs it from being distributed on resources limited cell phones. Further, dynamic analysis is liable for low-code coverage. Lately, the malware has been trying to recognize the emulator and other dynamic analysis framework and abstaining from exhibiting their payloads. Hence, some dynamic analysis frameworks are vulnerable to analysis evasion [29].

### 5. Hybrid Analysis

The hybrid analysis technique includes consolidating static and dynamic features gathered from examining the application and drawing data while the application is running, separately. Nevertheless, it would boost the accuracy of the identification [27], [30]. The principal drawback of hybrid analysis it consumes the Android system resources and takes a long time to perform the analysis.

## III. DEEP LEARNING ALGORITHMS

Adopting the right classification algorithm in line with the purpose of detection is essential considering its effects on detection performance and accuracy. In Table 1, the frequently used deep learning algorithms in Android malware detection are concisely introduced. For more information about deep learning algorithms, readers can refer to [35] and [38]. In [34] a detailed description of the variations of different deep learning algorithms is demonstrated. In addition [36], [37], and [38] shows the implementation of deep learning in different fields.

TABLE I COMMONLY USED DEEP LEARNING ALGORITHMS

| Algorithms | Strengths | Weaknesses |
|---|---|---|
| **Deep Neural Network (DNN)** common deep learning method used for classification. Comprised of more than 2 hidden layers. | Accomplished success in different applications | The learning process could be time consuming |
| **Restricted Boltzmann Machines (RBM)** are utilized as generative models of various kinds of data, which can study a likelihood dispersion over a specific arrangement of its inputs. In their conditional configuration, they can be utilized to show high dimensional temporal series, for example, sound or video streams. | - Allow to produce samples look as if they come from the data distribution<br>- It can be used as features extractor to train other models on top of it | - Hard to train well<br>- Computing the likelihood is time consuming |
| **Convolutional Neural Network (CNN)** A CNN comprises three layers, i.e., convolutional layer, subsampling layer (pooling layer) and fully-connected layer. The convolutional layer utilizes the convolution procedure to accomplish the weight sharing. The subsampling layer strives to decrease the dimension of the feature map. It can be applied by an average pooling procedure or a max pooling procedure. Thereafter, many fully-connected layers and a SoftMax layer are placed on the top layer for classification and recognition. The deep convolutional neural network commonly contains several convolutional layers and subsampling layers for feature learning on large-scale images. | - Fewer neuron connections needed in regard to a standard NN.<br>- Numerous variations to CNN have been developed | - Usually, it needs multiple layers to discover a complete hierarchy of visual features<br>- Commonly it needs a big dataset of tagged images |
| **Deep Belief Network (DBN)** made of different layers of stochastic, hidden variables. The upper two layers with indiscriminate, symmetric associations between them. The lower layers get top-down, directed associations from the layer above. | Offer layer by layer learning approach to initialize the network | The training phase consumes system resources because of the initialization process and sampling. |
| **Recurrent Neural Network (RNN)** is convenient to handle flows of data. They are made by one network doing a similar work for each component in a succession, with each yield value dependent on the past computations. | - Modeling time dependencies<br>- Able to remember serial events | Learning process suffers from vanishing gradient problem (large change in the value of parameters for the early layers doesn't have a big effect on the output) |
| **Deep Autoencoder is** a sort of DNN whose intended output is the data input itself. | - Applied to feature extraction/ dimensionality reduction.<br>- Many variants to DAE have been proposed | - It needs pre-training phase<br>- It Doesn't have the capacity to figure out what data is pertinent |

IV. RELATED REVIEWS

To our knowledge, this is the first review on the use of deep learning in Android malware detection. Notwithstanding, there exist reviews that surveyed different issues related to Android malware detection. K. Tam et al. [54] exhibited a wide review of Android malware analysis and detection techniques and evaluated their efficiency against progressive malware. Li Li et al. [48] presented a review of studies that statically analyze Android applications, from which they point out the directions of static analysis approaches, and reciting the major areas where future research is still needed. P. et al. [49] provided an extensive review of the works in dynamic mobile malware detection along with adequate analysis and comparison of the different approaches. Saba Rashid et al. [50] surveyed the work of hybrid Android malware analysis in a period of seven years. Then, they relied on the advantages and disadvantages of existing techniques to propose SAMADROID a hybrid Android malware analysis system. Alireza Sadeghi et al. [53] conducted an exhaustive review of Android security and came up with taxonomy to classify and identify the current research in this field.

We believe this work complements the previous reviews by surveying the use of deep learning in Android malware detection and fulfilling some research gaps in the field of Android malware analysis.

## V. THE METHODOLOGY

The methodology that applied to collect Data for this review
- Determination of the relevant information gathered from publications in the literature
- Then, listing the various search keywords that let the probing of the pertinent set of publications
- Conducting the search process in the publication repositories
- To confine the review on the pertinent papers, an exclusion criterion is implemented on the search results
- Finally, the papers are collected from different repositories to make the inclusive list of the review

### 1. Search keywords

The key word used to perform the search can be summarized as:

Deep learning for Android malware detection; applying /using deep learning techniques in Android malware detection + analysis. We also use some of the deep learning strategies in the search such as: using the convolutional neural network in Android malware analysis; using deep belief network in Android malware analysis; using the recurrent neural network in Android malware analysis.

### 2. The Search Databases

Repository probe is aimed at finding significant publications. The following online repositories are searched:

ACM Digital Libraries (https://dl.acm.org); Science Direct (https://www.sciencedirect.com); Web of Knowledge (https:// webofknowledge.com); IEEE Xplore Digital Library (https://ieeexplore.ieee.org); Cornell University Library (https://arxiv.org); SpringerLink (https://link.springer.com).

### 3 Exclusion Criteria

To make sure the search results are relevant to the review, the following exclusion criteria are used:
- Duplicated papers that are found in multiple repositories
- There are some papers about using Deep learning for malware detection in windows operating system, and in intrusion detection systems. All non-Android papers are removed
- Papers published in a non-English language
- Papers that are not published in their final version

In total, we collected 26 research papers from August 2014 (Date of publishing first work applied deep learning method in Android malware detection) to August 2018. 17 papers out of the 25 papers used static analysis. 4 papers employed dynamic analysis, and 5 papers utilized hybrid analysis.

## VI. LITERATURE REVIEW

### 1. Static Analysis

W. Li et al. [4] implemented a malware identification system utilizing deep learning method, which uses both dangerous API calls and risky permission combinations as features to construct a DBN model, which can automatically recognize malware from benign ones. They proposed scheme tested on the Drebin dataset, and the results showed that the model obtained 90% accuracy. Yi Zhang et al. [17] developed DeepClassifyDroid to identify Android malware resting on CNN. The proposed system executes static analysis to attain five diverse features. The system tested on a dataset contains 10,770 apps in total, involving 5546 malicious and 5224 benign. The results revealed that the approach's accuracy was 97.4%. M. Ganesh et al. [26] designed a deep learning system for Android malware detection. The proposed approach employs static analysis to acquire permission from the apps. Then, the requested permissions are converted into an image file that can be dealt with by a deep learning model. After that, the model is trained with image files. The model was tested on unbalanced data, including 2000 malware apps and 500 benign apps. The model accomplished 93% accuracy. The authors suggested that deep learning offers an extensible and exact solution for Android malware characterization because it determines malware relying on patterns instead of signatures.

Dali Zhu et al. [9] presented DeepFlow, a malware detection system that builds on data streams inside malignant apps that may contrast essentially from ones inside good apps, however, might be like different malignant apps to some degree. DeepFlow uses such contrasts and correspondences to consequently distinguish novel apps whether they are malignant or not by utilizing a deep learning model. Then, a classification model was built based on DBN. The proposed model was tested on 3000 benign apps and 8000 malicious apps and reached a 95.05% F1-Score. S. Hou et al. [12] proposed DroidDelver an Android malware detector stands on the API call block features drawn from the smali code. The extracted feature depicts statically drawn API call blocks from the smali codes. A DBN classifier was applied to classify unknown malware. DroidDelver evaluated on a dataset involved 2,500 benign apps and 2,500 and DroidDelver acquired 96.66% accuracy.

S. Hou et al. [5] suggested an alternative to applying API calls, they categorized the API calls which occurred in the same method within the smali code. Then, they developed AutoDroid (automatic Android malware detection) employing DBN depending on API call blocks. Experimental results performed on 2500 benign apps and 2500 malware showed that the best accuracy of the DBN was 95.98 %. L. Shiqi et al. [2] proposed image texture-based classification by converting APK binary data into image and utilized DBN to extract image texture. They also extracted some other features such as API calls, used permissions, and Activity. The scheme was tested on DREBIN dataset, and the results demonstrated that image texture when combined with API calls, gives the best accuracy of 95.6%. N. McLaughlin et al. [10] taking a different approach to facing Android malware challenges. They explored the utilization of CNN to malware identification by handling the dismantled byte-code of an application as a text to be broken down. This methodology has the benefit that features are automatically gained from raw data and subsequently eliminates the requirement for malware signatures to be planned by hand. The researchers performed experiments using three datasets: Genome project dataset; two additional large datasets provided by McAfee Labs. The model has different accuracy on different datasets, in the Genome dataset, the model has achieved 98% accuracy. In McAfee datasets, the model has accomplished 80% and 87% respectively.

R. Nix et al. [23] they concentrated on program analysis that looks at Android API calls made by an application. API calls depict how an application transmits information with the Android OS. Such transmission is basic for an application to do its jobs, thus giving essential data on an application's behavior. The authors planned a (CNN) sequence classification, which conducted convolution tasks along the sequences, learning successive shapes for every area as the convolution window slides down the sequence. The CNN structure additionally utilized different CNN layers to develop more elevated features from small local features. The model evaluated using a dataset includes 1016 APK files and obtained 99.4% accuracy. Wei Wang et al. [22] seeking to ameliorate accuracy of large-scale Android malware detection by designing a hybrid system formed on deep autoencoder (DAE) as pre-training procedure and different CNN structures for malware identification. An empirical test conducted on a dataset incorporates 23,000 apps, and results showed that the CNN-P structure gets a high accuracy of 99.8%.

E. Karbab et al. [3] developed MalDozer, a system depends on an artificial neural network that receives, as input, the raw sequences of API method calls, as they come from the DEX file, to allow malware identification and family ascription. In training, MalDozer can determine malicious patterns in an automatic way utilizing just sequences of raw method calls in the assembly code. They performed a comprehensive test on various datasets containing malware and good apps. The results exhibited that MalDozer was productive and viable in malware detection, but less effective in family attribution.

T. Huang and H. Kao [11] proposed a color-compounded convolutional neural networks-based Android Malware Detection that instead of features extraction depending on color representation for transforming Android apps into RGB color code and converting them to unchanged-sized translated images. Afterward, the encoded image is sent to CNN to automate feature extraction and learning. The proposed approach tested on huge dataset, the best accuracy achieved, was 93%. C. Hasegawa and H. Iyatomi [1] developed a light-weight malware identification system that utilized 1-D CNN and analyzed a small part of the raw APK file without unpackaging. A test conducted on dataset consists of 5,000 malware and 2,000 good apps. Although high accuracy was reported, a further evaluation of the suggested approach is needed.

K. Xu et al. [25] suggested the DeepRefiner malware characterization system uses deep neural networks with different hidden layers to automate features extraction. In a preprocessing step, DeepRefiner recovered XML values from XML files in the first detection layer and seized bytecode semantics from the dismantled classes.dex file in the second detection layer. DeepRefiner then signified apps as vectors, which were utilized as inputs for deep neural networks. The hidden layers in neural systems consequently build identification features from input vectors through non-linear translation. DeepRefiner was tested on a large dataset, including 62,915 malware apps and 47,525 benign apps. The results showed that DeepRefiner malware detection accuracy was 97.74%. D. Li et al [6] Developed a system built on a Deep neural network that applied static features namely required permissions and API calls to train the model to classify Android malware. DREBIN dataset was used to evaluate the performance of the model which reached 95.64% F1-Score.

X. Su et al. [13] designed DroidDeep a deep learning strategy for Android malware identification which examined different levels of features. DroidDeep first explored static data involving permissions, API calls, and deployment of components for distinguishing the behavioral patterns of Android apps and drawing out multiple features set from Android apps. Then, these features were fed into a deep learning model to study representative features for classification. Lastly, the studied features were forwarded into a detector built on the Support Vector Machine (SVM) for identifying Android malware. DroidDeep was tested with 3,986 benign apps and 3,986 malware and accomplished 99.4% accuracy. Zi Wang et al. [21] designed DroidDeepLearner, a malware identification system for Android platform employing deep learning an algorithm, which utilizes both dangerous API calls and risky permission combinations as features to construct a DBN model, which is capable of automatic detection of Android malicious apps from benign ones. Experiments carried out on 6,334 apps in total and DroidDeepLearner achieved 93.09% accuracy.

2. Dynamic Analysis

H. Liang et al. [16] developed natural language processing techniques for Android malware analysis on the assumption that there is a similarity between theme drawing and malware identification. They designed a model that deals with system call sequences as texts and considers the malware detection function as theme extraction. The proposed approach was tested on a dataset of 14,231 apps. The results showed that the accuracy was 93.16%. F. Martinelli et al. [7] established their work on the hypothesis that there is a correspondence between sentiment analysis and malware analysis. Negative and positive sentiments are equivalent to maliciousness and benignity of apps. They built a deep learning model that used CNN to distinguish between trusted and malicious applications. The model tested on a dataset contains 7,100 apps, and obtained accuracy between 85% and 90%.

S. Hou et al. [18] proposed an automatic Android malware detection system Deep4MalDroid. Depending on a graph representation of extracted Linux system call, the Stacked Autoencoders (SAEs) was applied to scan general patterns of malware and thus to determine newly unknown malicious applications. Experimental results on a dataset of 3000 apps showed that the best accuracy reached was 93.68%. L. Singh and M. Hofmann [20] mainly developed conventional machine learning models and one deep learning model standing on runtime behavior (system calls). Experimental results showed that the accuracy of 97.16% was achieved using the SVM classifier, which performed better than Decision Tree, Random Forest, Gradient boosted trees, Neural network, K-NN, and DNN. This is because the utilized data set was small, it contains 494 malicious and benign apps. As deep learning is data hungry testing with larger dataset can give different results.

3. Hybrid Analysis

Z. Yuan at al. developed Droid-sec [8] considered the first attempt to apply deep learning in Android malware detection. The authors use numerous features to compose Deep Belief network models that are able to classify malware from benign ones. The model performance evaluated on small dataset includes 250 benign apps and 250 malware samples. The results emphasized the use of deep learning in Android malware identification, and the model obtained 96% accuracy. Z. Yuan at al. Designed Droid Detector [24] another model based on DBN. The proposed approach was tested on a large unbalanced dataset that involves 20000 benign and malware samples. The results indicated a good performance of DBN with 96.76% accuracy. L. Xu et al. presented HADM [14] an Android malware identification approach that depends on autoencoders to study the features of the apps, then built an SVM classifier to differentiate the apps as malicious or trusty. They conducted experiments on a dataset of 5888 benign and malware and evaluated the performance of static and dynamic features separately. The result showed that static features have better performance than dynamic features.

H. Alshahrani et al. [15] developed DDefender a malware identification system comprises of two fundamental parts: First, client side, a light application running on the client's phone to preform dynamic analysis and present clients with analysis results. Second, server side, a framework that preforms static analysis and detection procedure and sends the outcomes back to the client side. The dynamic analysis was used to draw system calls, system information and network traffic from targeted applications. Then static analysis employed to draw substantial features from the targeted application. A deep neural network applied to build a mode that tested on a dataset of 4208 benign and malicious apps. The model achieved 95% accuracy. R. Vinayakumar et al [19] proposed the use of Long Short-Term Memory (LSTM) which is a particular kind of recurrent neural network applied to study long-term transient dynamics with a series of random lengths for Android malware characterization. The authors extracted dynamic and static features and used a dataset of 1738 for model performance evaluation, which accomplished 93.9% in dynamic analysis and 97.5 in static analysis.

Premised on the previous review we compare these approaches using 10 criteria separately in Tables 2-4. These criteria are: key concept; features used for malware detection; dataset; place of analysis (on device, server, etc...); Realtime detection; algorithms used; used measures with their values; contribution; limitation; availability for public (check if the researchers allow access to their dataset and work for the public).

VII. CHALLENGES, OPEN ISSUES, AND FUTURE DIRECTIONS

1. Open Issues

According to the review and discussion, in the following sections, a number of open issues are presented.

1.1 Datasets Issues

There are some problems regarding the datasets used for model evaluation such as:

Getting Representative data set that reflect the distribution of malicious and benign apps in the real world. How to determine and select strong, discriminative features whether they are hand engineered or selected by an algorithm persists open issue. Moreover, as shown in Tables 2-4 there are only two works that make their work accessible to the public. Lack of standard Android malware datasets called for sharing datasets among research communities.

TABLE 2 COMPARISON OF ANDROID STATIC ANALYSIS MALWARE DETECTION TECHNIQUES

| Ref. | Publishing Year | Key Concept | Features | Dataset | Total Dataset | Place of Analysis | Real Time Detection | Used Algorithms | Measures | Values | Contribution | Limitation | Availability to the public |
|---|---|---|---|---|---|---|---|---|---|---|---|---|---|
| [13] | 2016 | Static Features | Requested permissions-used permissions-sensitive API calls-Actions-app components | 3986 B from Google play-3986 M from DREBIN | 7972 Apps | Computer | No | DBN | F1 Prec. Recall | 97.3 98.2 98.4 | Presented DroidDeep for malware detection using DBN | Susceptible to adversarial attack | No |
| [21] | 2016 | Static Features | Risky Permissions-dangerous API calls | - | 6334 Apps | Computer | No | DBN | F1 Prec. Recall | 94.5 93.09 94.5 | Proposed DroidDeepLearner combine risky permission and dangerous API calls to build a DBN classification model. | Susceptible to adversarial attack | No |
| [12] | 2016 | Static Features | API call blocks | Comodo cloud security- 2500 B-2500 M | 5000 Apps | Computer | No | DBN | ACC | 96.66 | Developed DroidDelver A detection system utilizes an API call block to identify malware. | Susceptible to adversarial attack | No |
| [26] | 2017 | Static Features | Requested permission | 5000 B from APKMirror-APK4fun 2000 M from Genome-DREBIN | 2500 Apps | Server | Yes | CNN-AlexNet | Acc | 93 | Proposed a detection system that transforms requested permissions into an image format, then use CNN for classification | - Depend only on permissions with their known limitations - Susceptible to impersonate attack | No |
| [9] | 2017 | Static data flow analysis | 323 features | 3000 B from Google play-8000 M from virusshare and Genome | 11000 Apps | Computer | No | DBN | F1 | 95.05 | Designed FlowDroid an identification system leverages data flow analysis to identify malware. | Susceptible to adversarial attack | No |
| [5] | 2017 | API calls | 1058 API calls | Comodo cloud security-2500 B-2500 M | 5000 Apps | Computer | No | DBN SAE | ACC ACC | 96.66 95.98 | Developed AutoDroid for malware detection using DBN and SAE | Susceptible to adversarial attack | No |
| [10] | 2017 | Opcode Sequence | Learn to detect sequences of opcode that indicate malware | Genome-McAfee Labs | 27377 Apps | Computer | No | CNN | ACC Prec. Recall F1 | 98 99 95 97 | Developed a detection system that relies on automatic features learning from raw data and treating the disassembled code as text | Although trained on a large dataset, performance dropped when tested on a new dataset- Susceptible to impersonate attack | No |
| [23] | 2017 | Static Features | API call sequence | 216 M from Contagio minidump-1016 B from third-party market | 1232 Apps | Computer | No | CNN | Acc Prec. Recall | 99.4 100 98.3 | Apply multiple layers of CNN to learn the features, then classify apps based on API call sequence | Susceptible to impersonate attack | No |

TABLE 2 (continued)

| Ref. | Publishing Year | Key Concept | Features | Dataset | Total Dataset | Place of Analysis | Real Time Detection | Used Algorithms | Measures | Values | Contribution | Limitation | Availability to the public |
|---|---|---|---|---|---|---|---|---|---|---|---|---|---|
| [11] | 2017 | Transfer classes.dex into RGB color images | Extract features from the transferred images | Collected by the researchers | 829356 Apps | Server | No | CNN | Acc | 97.7 | Proposed color representation for translating Apps into RGB color image which is fed to CNN for automatic features learning | Results revealed that human experts are still needed in long-term sample collection and model updates. Susceptible to impersonate attack | Shared the converted images on a website |
| [4] | 2018 | Static features | Dangerous API calls-risky permissions | 1400 B from Google play-1400M from DREBIN | 2800 Apps | Computer | No | DBN | Recall | 94.28 | Utilized DBN to build automatic malware classifier | Susceptible to adversarial attack | No |
| [17] | 2018 | Static features | API calls-Permissions-Intent filters | 5224 B from Chinese third-party market-5546 M from DREBIN | 10770 Apps | Computer | No | CNN | Prec Recall ACC F1 | 96.6 98.3 97.4 97.4 | Presented DeepClassifyDroid Android malware detection system based on CNN | Susceptible to impersonate attack | No |
| [2] | 2018 | Static features | API calls | DREBIN | 6965 Apps | Computer | No | DBN | Acc | 95.7 | Proposed an image texture analysis approach for malware detection | Susceptible to adversarial attack | No |
| [22] | 2018 | Static features | Permissions-requested permissions-filtered intents-restricted API calls-hardware features-code related features-suspicious API calls | 10000 B from Anzhi play store-1300 M from virusshare | 23000 Apps | Computer | No | CNN | Acc Recall F1 | 99.8 99.91 99.82 | Developed a hybrid model for malware detection using CNN and DAE | Susceptible to impersonate attack | No |
| [3] | 2018 | Static features | API sequence calls | Genome-Virusshare-DREBIN-Contagio minidump | 33000 Apps | IOT devices-Computer-Server | Yes | CNN | F1 Prec Recall | 96.29 96.29 96.29 | Designed MalDozer an Android malware detection system that identifies malware and tries to attribute it to its family by applying NLP techniques | - Affected by dynamic code loading and obfuscation. Besides, family attribution is limited - Susceptible to impersonate attack | No |

TABLE 2 (continued)

| | Publishing Year | Key Concept | Features | Dataset | Total Dataset | Place of Analysis | Real Time Detection | Used Algorithms | Measures | Values | Contribution | Limitation | Availability to the public |
|---|---|---|---|---|---|---|---|---|---|---|---|---|---|
| [25] | 2018 | Static Features | The semantic structure of Android bytecode | Google play-virusshare-Masset | 110440 Apps | Computer | No | CNN-LSTM | Acc | 97.74 | Proposed DeepRefiner to identify malware by using LSTM on the semantic structure of Android bytecode | It needs a frequent update with new labeled features- Heavy computation resources are required- Susceptible to impersonate attack | No |
| [6] | 2018 | Static Features | Permissions- API Calls | DREBIN – 5560 M- 123453 B | 129013 Apps | Computer | No | DNN | Prec Recall F1 | 97.15 94.18 95.64 | Implemented malware detection engine based on DNN | Susceptible to adversarial attack | No |
| [1] | 2018 | Static Features | Code Analysis | 2000 B from APKpure and APPsapk- 5000 M from AMD and DREBIN | 7000 Apps | Computer | No | CNN | Acc | 95.4 | Proposed a method that uses 1-D CNN to analyze a small portion of raw APK | The approach requires further analysis to prove it is efficien | No |

TABLE 3 COMPARISON OF ANDROID DYNAMIC ANALYSIS MALWARE DETECTION TECHNIQUES

| Ref. | Publishing Year | Key Concept | Features | Dataset | Total Dataset | Place of Analysis | Real Time Detection | Used Algorithms | Measures | Values | Contribution | Limitation | Availability to the public |
|---|---|---|---|---|---|---|---|---|---|---|---|---|---|
| [18] | 2016 | Dynamic Features | System calls | Comodo cloud security- 1500 B- 1500 M | 3000 Apps | Computer | No | SAE | Acc | 93.68 | Proposed dynamic behavior malware detection by monitoring the system call of an app during execution | The proposed methodology may not cover all events during execution. Besides, it is not immune to evasion techniques | No |
| [7] | 2017 | Dynamic Features | System Calls | 3536 B from Google Play – 3564 from DREBIN | 7100 Apps | Computer | No | CNN | Acc | 85-95 | Developed a detection system built on CNN that utilizes system calls collected from the dynamic run of apps | Apps are executed for 60 seconds only, which is not enough to capture all app events. Additionally, the method is not immune to evasion techniques | No |
| [16] | 2017 | Dynamic Features | System Call Sequences | 10000 B, 4231 M | 14231 Apps | Computer | No | CNN | Acc Prec F1 | 93.1 95.75 86.57 | Designed an end-to-end malware identification model by considering system call sequences as text and regarding the malware characterization as theme extraction | The approach is not resistant to evasion techniques | No |

*10*

TABLE 4 COMPARISON OF ANDROID HYBRID ANALYSIS MALWARE DETECTION TECHNIQUES

| Ref. | Publishing Year | Key Concept | Features | Dataset | Total Dataset | Place of Analysis | Real Time Detection | Used Algorithms | Measures | Values | Contribution | Limitation | Availability to the public |
|---|---|---|---|---|---|---|---|---|---|---|---|---|---|
| [8] | 2014 | Dynamic/Static features | Sensitive API calls | 250 B from google play-250 M from contagion minidump | 500 Apps | Computer | No | DBN | Acc | 93.5 | The first paper examined the use of deep learning in Android malware detection | The proposed scheme was evaluated on a small dataset. | Yes at the time of publishing |
| [24] | 2016 | Dynamic/Static features | Required permissions-sensitive API calls- apps actions collected dynamically | 20000 B from Google Play – 17604 from Genome and contagion minidump | 21700 Apps | Server | Yes | DBN | Prec Recall | 97.79 95.68 | Developed a system that aggregates static and dynamic features and applies DBN to detect malware | The dynamic analysis may not detect malware provided with anti-analysis techniques | No |
| [14] | 2016 | Dynamic/Static features | 9 static features-system call from the dynamic running of an app | 4002 B from Google play, 1886 M from virusshare | 5888 Apps | Computer | No | DNN | Acc | 93.4 | Designed HADM a hybrid Android malware identification system based on DNN | Malware can detect the emulator used in dynamic analysis and may not exhibit any malicious behavior | No |
| [15] | 2018 | Dynamic/Static features | Permissions-Intents-app components-network activities-Linux system call- | 2104 B, 2104 M from DREBIN and Marvin | 4208 | Server | Yes | DNN | ACC | 95 | Developed DDefender Android malware identifier that can run on user's device | Monkey Tool used in the dynamic analysis may not be able to generate all the events that a malware can make. There is an overhead for uploading an APK to a server for analysis. | No |
| [19] | 2018 | Dynamic/Static features | Permissions-events generated by Monkey Tool | 279 B and 279 M for static analysis-408 B and 1330 M for dynamic analysis | 2296 | Computer | No | RNN LSTM | | | Proposed use of RNN and LSTM for Android malware detection | Apps were running in the emulator for a short time. Malware can detect there are being run in the emulator and refrained from showing any malicious activities | No |

B: benign. M: malware. Acc: accuracy. Prec: precision. F1: F1-Score. CNN: convolutional neural network. SAE: Stacked Autoencoder. LSTM: Long Short-Term Memory. RNN: recurrent neural network. DNN: deep neural network.

*11*

1.2 Concept Drift

Concept drift is a problem that ensues the swift rise in the number of Android malicious apps and trusty apps along with fast technological development in the Android environment [43]. As can be seen in Tables 2-4 old datasets such as DREBIN are still widely used, and this makes the models built on it liable for concept drift. DeepRefiner [25] is the only work among the reviewed papers that implicitly approached the concept drift problem and suggested continual updates of the model. Otherwise, the model will yield equivocal predictions. How to effectively discern the existence of concept drift and solve it stays an open issue.

1.3 Deep Learning Security Issues

There are challenges regarding attacking deep learning models (including Android malware detection models) in the training phase and testing phase abided unsolved. In the training phase, the models are subject to data poisoning attacks, which are merely implemented by manipulating the training and instilling data that make a deep learning model to commit errors. In the testing phase, the models are exposed to several attack types [44]:

- Adversarial Attacks: are attacks into deep learning model inputs that an adversary has invented deliberately to cause the model to make mistakes,
- Evasion attack: the intruder exploits malevolent instances at test time to have them incorrectly classified as benign by a trained classifier, without having an impact over the training data. The intruder's objective in this manner adds up to breaching system integrity, either with a targeted or with an indiscriminate attack according to his purpose,
- Impersonate attack: opt to mimic data instances from targets, distinctly, an attacker plans to create particular adversarial instances to such an extent that current deep learning-based models mistakenly characterize original instances with different tags from the imitated ones,
- Inversion attack: use the APIs allowed by machine learning systems to assemble some fundamental data with respect to the target system models. This kind of attack is divided into two types: white-box attack and black-box one. In particular, the white-box attack implies that an aggressor can loosely get to and download learning models and other supporting data, while the black-box one points to the way that the aggressor just knows the APIs opened by learning models and some observation after providing input.

All of the reviewed works could be vulnerable to one or more types of these attacks. Amongst the reviewed works, only DeepRefiner [25] encountered adversarial sampling and attempted adversarially training the model. No sufficient results were reported, the researchers planned to harden their model in future work. Hardening deep learning models against adversarial attacks is one of the greatest challenges facing deep learning methods.

1.4 Deep Learning Issues

Recently deep learning has been subjected to a rigid assessment from different views such as [42], [46], and [41]. Here we reflect the open issues related to the use of deep learning in Android malware detection:

- Currently available machine learning has several weaknesses such as there are all can be deceived,
- Deep learning lacks transparency to provide an interpretation of the decision created by its methods. Malware analysts need to understand how the decision was made,
- there is no assurance that classification models built based on deep learning will perform in different conditions with new data that would not match previous training data,
- Deep learning study complex correlations within input and output feature with no innate depiction of causality,
- Deep learning models are not autonomous and need continual retraining and rigorous parameters adjustments.

As shown in Tables 2-4 there are works such as [10] and [16] tried to automate feature engineering and model updates, results demonstrate that deep learning models up to now are not autonomous, human expert involvement in feature engineering and model updates is irreplaceable. Additionally, as found in [11] there is no guarantee that deep learning models will be efficient in the new dataset that might not match previously trained dataset.

1.5 Data Privacy Preservation

Some work for instance [11] and [26] sends data for analysis to a remote server, in such a situation the privacy of the data ought to be preserved. Despite a previous study [49] have referred to this issue, it has stayed unaddressed.

2. Future Directions

The objective of analysis: according to the taxonomy of [53], the objectives of Android malware analysis is divided into three: malware detection, vulnerability detection, and grayware detection. As can be seen from Tables 2-4 the malware detection dominated the work. In the future more, work to explore vulnerability and grayware detection with deep learning is needed.

2.1 Static Analysis

As reported in [53] the static analysis contains data structure analysis and sensitivity analysis. Further data structure analysis divided into text-based, control flow graph (CFG), call graph (CG), and inter-procedural control flow graph (ICFG). Sensitivity analysis is divided into flow sensitive, context sensitive, and path sensitive. As shown in Table 2 only data flow analysis has been used with deep learning methods. Future works in static analysis may examine the use of deep learning with these techniques.

### 2.2 Dynamic Analysis

In line with [53] taxonomy, there are two aspects of dynamic analysis inspection level and input generation. Inspection levels include application levels that trace java method invocation; kernel level collects system calls using kernel modules; the virtual machine intercepts events that occur within emulators. Obviously, from Table 3 deep learning has been applied to all inspection level methods. On the other hand, input generation involves fuzz testing using Monkey, which is largely adopted in dynamic analysis, and symbolic execution which depends on symbolic values instead of actual values for program input. None of the reviewed work utilized dynamic analysis applied symbolic execution, future work in dynamic analysis or hybrid analysis may consider taking advantage of symbolic execution.

### 2.3 Deep Learning Security

More studies are required to resolve the issues discussed in Section 7.1.4. Albeit the efforts to tackle these issues as mentioned in [44], still, there are no satisfactory solutions. Possible future directions are:

- Improving the Robustness of the Learning Algorithms: by extending existing algorithms and examining the effectiveness of concepts such as stability training and smoothing model output,
- Designing secure algorithms: to build algorithms that can defend against evasion attacks,
- Apply dimension reduction strategy: this strategy aspires to improve the flexibility of a classifier by lowering the dimension of features.

### 2.4 Hardening Deep Learning Models Against Adversarial Attacks

Researchers in [45], and [51] proposed retraining and distillation to confront adversarial attacks on deep learning models. The results show that there are no perfect solutions, partly because of these solutions originally developed for fighting adversarial attacks in the computer vision field, which is different from malware detection. More studies are required to investigate the use of the following techniques:

- Distillation: the notion of distillation is that suppose there is already a neural network F that can classify training set x into target class y and generates as output a probability distribution over y. This output is utilized as tags to train a second model F'. Whereas the new tags include more information about the data x than the first model, the network will function similar to the first network,
- Retraining: to retrain the classifier with adversarial generated instances. The purpose of retraining is to amend the abstraction of the model. The challenge is hidden in selecting the proper number of adversarial training instances,

### 2.5 Data Privacy Preservation

Considering the utilization of data encryption techniques is one way to achieve the goal of the user's data privacy when uploading files for further analysis to a remote server.

### 2.6 Concept Drift

How to describe the concept's drift, and how to measure drift magnitude, is significant to solve the concept's drift problem. Ideas presented in [47] require further studies to verify if there are applicable to Android malware detection.

### 2.7 Performance Evaluation

In the reviewed research there are several works used unbalanced dataset for example [24] and [26], in such cases considering the use of phase response curve (PRC) as proven in [52] to be a better metric in displaying divergences between algorithms that are not obvious in ROC space.

## VIII. CONCLUSION

In this work, we presented a thorough review of the use of deep learning in Android malware detection. A comparison of existing work with respect to certain criteria was presented. The review uncovered knowledge gaps in the existing work and underscores major challenges and open issues that will direct future research efforts. The review showed that the current literature could be exposed to different adversarial attacks and subjected to concept drift. Static Android malware analysis dominates the existing work. Future work may consider dynamic research techniques or utilizing hybrid analysis techniques. Sharing research datasets and tools between researchers lingered unaddressed except in a few cases. Hardening deep learning models against different adversarial attacks and detecting, describing and measuring concept drift are vital in future work in Android malware detection. Furthermore, researchers need to bear in mind the limitation of deep learning methods such as lack of transparency and nonautonomous of it is model to build more efficient models. Finally, the results of this work can aid to promote research in Android malware detection based on deep learning methods.